\documentclass[12pt]{article}

\usepackage{graphicx}

\begin{document}
%\titlerunning{Title running}
\begin{center}
{\Large\bf \boldmath On Lagrangian formulations for mixed-symmetry
HS fields on AdS spaces within BFV-BRST approach} %<== title (bold face, capitalize)

\vspace*{6mm}
{A.A. Reshetnyak}\\      %<== authors
{\small \it Laboratory of Nonlinear Media Physics, Institute of Strength Physics and Materials Science SB of RAS, 634021, Tomsk, Russia}  %<== institutions

\end{center}

\vspace*{6mm}

% abstract
\begin{abstract}
The key aspects of a gauge-invariant Lagrangian description of
massive and massless half-integer higher-spin fields in AdS spaces
with a two-row Young tableaux $Y(s_1,s_2)$ are presented in an
unconstrained description, as well as in off-shell formulations
with algebraic constraints, on the basis of BFV-BRST operators for
non-linear operator superalgebras, encoding the initial conditions
realized by constraints in a Fock space and extracting the
higher-spin fields from unitary representations of the AdS group.
\end{abstract}

\vspace*{6mm}

Problems of a unified description of the known interactions and
the variety of elementary par\-tic\-les are revealed at high
energies (partially accessible in LHC), thus providing the
relevance of the development of higher-spin (HS) field theory due
to its close relation to superstring theory in constant-curvature
spaces, which operates with an infinite set of bosonic and
fer\-mi\-o\-nic HS fields subject to a multi-row Young tableaux
(YT) $Y(s_1,...,s_k),k\hspace{-0.3ex} \geq\hspace{-0.3ex}1$ (for a
re\-view, see \cite{reviews}). This report deals with the recent
results in constructing Lagrangian formulations (LFs) for free
fermionic HS fields in $AdS_d$-spaces with $Y(s_1,s_2)$ in a
Fronsdal metric-like formalism within the BFV-BRST approach
\cite{BFV}, as a starting tool for a description of interacting HS
fields in the framework of quantum field theory, and is partially
based on the results presented in \cite{adsfermBKR,flatfermmix}.

This method of constructing an LF for HS fields, originally
developed as applied to a Hamiltonian quantization of gauge
theories with a given LF, consists in a solution of the problem
\textit{inverse} to that of the method \cite{BFV} (as in the case
of string field theory \cite{SFT} and the first papers on HS
fields \cite{Ouvry}) in the sense of constructing a gauge LF
w.r.t. a nil\-po\-tent\nolinebreak{\,}BFV-BRST\nolinebreak{\,
}operator\nolinebreak{\ }$Q$. $Q$ is constructed from a system
$O_\alpha$ of 1st-class const\-raints, including a special
non-linear\nolinebreak{\ }ope\-ra\-tor su\-per\-al\-geb\-ra
$\{O_I\}$:$\{O_I\} \supset \{O_\alpha\}$, defined on an auxiliary
Fock space and encoding\nolinebreak{\,}the relations that extract
the fields with a fixed $(m,\mathbf{s})$ from the AdS-group
representation spaces.

A massive representation with spin $\mathbf{s}=(s_1,s_2)$, $s_i = n_i+\frac{1}{2}$,
$n_1 \geq n_2$, of the AdS group in an AdS$_d$ space
is characterized by $Y(s_1,s_2)$ and realized in the space of
mixed-symmetry spin-tensors %with suppressed Dirac index
\begin{equation}\label{Young k2}
\Phi_{(\mu)_{n_1},(\nu)_{n_2}} \equiv
\Phi_{\mu_1\ldots\mu_{n_1},\nu_1\ldots\nu_{n_2} A}(x)
\longleftrightarrow
\begin{array}{|c|c|c c c|c|c|c|c|c| c| c|}\hline%\vphantom{\biggm|}
  \!\mu_1 \!&\! \mu_2\! & \cdot \ & \cdot \ & \cdot \ & \cdot\ & \cdot\ & \cdot\  & \cdot\ &
  \cdot\
  & \cdot\    &\!\! \mu_{n_1}\!\! \\
   \hline%\vphantom{\biggm|}
    \! \nu_1\! &\! \nu_2\! & \cdot\
   & \cdot\ & \cdot & \cdot & \cdot & \cdot & \cdot & \!\!\nu_{n_2}\!\!   \\
  \cline{1-10}%\vphantom{\biggm|}
\end{array}\ ,
\end{equation}
subject to the equations (for $\beta = (2;3) \Longleftrightarrow
(n_1>n_2; n_1 = n_2$), $r$ being the inverse squared AdS$_d$-radius,
with a suppressed Dirac index $A$ and the matrices
$\bigl\{\gamma_\mu,\gamma_\nu\bigr\}=2g_{\mu\nu}(x))$,
\begin{eqnarray}
\label{Eq-0} \left(\bigl[i\gamma^{\mu}\nabla_{\mu}
    -r^\frac{1}{2}(n_1  + \textstyle\frac{d}{2}-\beta)-m
\bigr],\ \gamma^{\mu_1} ,\ \gamma^{\nu_1}
\right)\Phi_{(\mu)_{n_1},\ (\nu)_{n_2}} =
\Phi_{\{(\mu)_{n_1},\nu_1\}\nu_2...\nu_{n_2}}=0.
\end{eqnarray}
To describe simultaneously all fermionic HS fields, one introduces
a Fock space $\mathcal{H}=\mathcal{H}^1\otimes\mathcal{H}^2$
generated by 2 pairs of creation $a^i_\mu(x)$ and annihilation
$a^{j+}_\mu(x)$ operators,
$i,j\hspace{-0.15ex}=\hspace{-0.15ex}1,2,
\mu,\nu\hspace{-0.15ex}=\hspace{-0.15ex}0,1...,d-1$: $[a^i_\mu,
a_\nu^{j+}]\hspace{-0.15ex}=\hspace{-0.15ex}-g_{\mu\nu}\delta_{ij}$,
and a set of constraints for an\nolinebreak{\,}arbitrary
string-like\nolinebreak{\,}vector\nolinebreak{\,}$|\Phi\rangle$:
\begin{eqnarray}
\label{t't0} \hspace{-1ex} {\tilde{t}}'_0|\Phi\rangle &
\hspace{-1ex}= \hspace{-1ex}& \bigl[-i\tilde{\gamma}^\mu D_\mu +
\tilde{\gamma}\bigl( m + \sqrt{r} (g^1_0 -
\beta)\bigr)\bigr]|\Phi\rangle=0 ,\quad \bigl({t}^i,
t\bigr)|\Phi\rangle = \bigl(\tilde{\gamma}^\mu a^i_\mu, a^{1+}_\mu
a^{2\mu}\bigr) |\Phi\rangle=0,\\
\label{PhysState} \hspace{-1ex} |\Phi\rangle & \hspace{-1ex}=
\hspace{-1ex}&
\textstyle\sum_{n_1=0}^{\infty}\sum_{n_2=0}^{n_1}\Phi_{(\mu)_{n_1},(\nu)_{n_2}}(x)\,
a^{+\mu_1}_1\ldots\,a^{+\mu_{n_1}}_1a^{+\nu_1}_2\ldots\,a^{+\nu_{n_2}}_2|0\rangle,\
|\Phi\rangle \in \mathcal{H},
\end{eqnarray}
equivalent to Eqs. (\ref{Eq-0}) for all $\mathbf{s}$, and given in
terms of an operator $D_\mu$ equivalent to $\nabla_{\mu}$,
in its action on $\mathcal{H}$. The fermionic operators
${\tilde{t}}'_0, {t}^i$ are defined through a set of
Grassmann-odd gamma-matrix-like objects, $\tilde{\gamma}^\mu,
\tilde{\gamma}$ ($\{\tilde{\gamma}^\mu,\tilde{\gamma}^\nu\} =
2g^{\mu\nu}$, $\{\tilde{\gamma}^\mu,\tilde{\gamma}\}=0$,
$\tilde{\gamma}^2=-1$ \cite{adsfermBKR}), related to the conventional
gamma-matrices by an odd non-degenerate transformation: $\gamma^{\mu}
= \tilde{\gamma}^{\mu} \tilde{\gamma}$.
\newline\indent
To derive a Hermitian BFV-BRST charge $Q$, whose cohomology in
zero ghost number subspace of a total Hilbert space
$\mathcal{H}_{\mathrm{tot}} = \mathcal{H}\otimes
\mathcal{H}'\otimes \mathcal{H}_{\mathrm{gh}}$ will coincide with
space of solutions for Eqs. (\ref{Eq-0}), we need to deduce a set
of 1st-class quantities, $O_I$: $\{O_\alpha\}\subset \{O_I\}$,
closed under the Hermitian conjugation w.r.t. an odd scalar
product $\langle\tilde{\Psi}|\Phi\rangle$ \cite{flatfermmix}, with
the measure $d^dx \sqrt{-{\mathrm{det}}g}$ and supercommutator
multiplication $[\ ,\ \}$.\hspace{-0.15ex} As a result, the
massive half-integer HS symmetry superal\-geb\-ra in $AdS_d$
spaces with $Y(s_1,s_2)$:
$\mathcal{A}(\hspace{-0.1em}Y(2),\hspace{-0.1em}AdS_d\hspace{-0.1em})$=$\{o_I\}$=$\{{\tilde{t}}'_0,{t}_i,{t}_i^{+},
t$, $t^+,{l}_i,{l}_i^{+},{l}_{ij}, {l}_{ij}^+,
g_0^i,\tilde{l}_0'\}$,
\begin{equation}
 \bigl(t^{i+};g^i_0; t^+;l^i, l^{+i};l_{ij}\bigr) \hspace{-0.2ex}=\hspace{-0.2ex}
  \bigl(\tilde{\gamma}^\mu a_\mu^{i+};
-a^{i+}_\mu a^{\mu{}i}+{\textstyle\frac{d}{2}}; a^{\mu{}1}
a_\mu^{2+};-i(a^{\mu{}i}, a^{+\mu{}i})D_\mu;
{\textstyle\frac{1}{2}}a^\mu_i a_{\mu{}j}\bigr),\ i\leq j
%\label{l2+}
\end{equation}
\vspace{-2em}
\begin{eqnarray}
{\tilde{l}}'_0=g^{\mu\nu}(D_\nu
D_\mu-\Gamma^\sigma_{\mu\nu}D_\sigma)
-r\Bigl(\sum\nolimits_i(g^i_0+t^{i+}t^i)+{\textstyle\frac{d(d-5)}{4}}\Bigr)
+ \Bigl(m + \sqrt{r} (g^1_0 - \beta)\Bigr)^2, \label{l'l0}
\end{eqnarray}
contains a central charge $\tilde{m}=(m-\beta\sqrt{r})$, a subset
of (4+12) differential $\{l_i, l_i^+\} \subset \{o_\mathbf{a}\}$
and algebraic $ \{t_i,t^+_i, t, t^+, l_{ij}, l_{ij}^+\} \subset
\{o_\mathbf{a}\} $ 2nd-class constraints, and particle-number
operators, $g_0^i$, composing, together with $\tilde{m}$, an
invertible supermatrix $\|[o_\mathbf{a}, o_\mathbf{b}\}\| =
\|\Delta_{\mathbf{ab}}(g_0^i,\tilde{m})\| +
\|\mathcal{O}(o_{I})\|$.

Among the 2 variants of an additive conversion for non-linear
superalgebras \cite{0711.4489SQS07} of $\{o_I\}$ into a 1st-class
system $\{O_\alpha\}$, 1) $\{o_\mathbf{a}\}$ results in an
unconstrained LF; 2) the differential constraints and a part of
the algebraic ones, $l_i, l_i^+,t, t^+$, restrict
$\mathcal{A}(Y(2),AdS_d)$ to the surface $\{o^r_\mathbf{a}\}\equiv
\{o_\mathbf{a}\}\setminus\{l_i$, $l_i^+, t, t^+\}$ on every stage
of the construction, resulting in an LF with off-shell
\-tra\-ce\-less and $\gamma$-\-tra\-ce\-less conditions, we
consider in detail the first case. To find the additional parts
$o'_I$: $o_I \hspace{-0.1ex}\rightarrow
\hspace{-0.1ex}O_I\hspace{-0.1ex}$ $=\hspace{-0.1ex} o_I
\hspace{-0.1ex}+ \hspace{-0.1ex}o'_I,
[o_I,o'_I\}\hspace{-0.15ex}=\hspace{-0.15ex}0$ such that $[O_I,
O_J\}\hspace{-0.1ex} \sim \hspace{-0.1ex}O_K$, we need: a)
following \cite{adsfermBKR, 0711.4489SQS07}, to pass to another
basis of constraints, $o_I\hspace{-0.15ex} \rightarrow
\hspace{-0.15ex}\tilde{o}_I \hspace{-0.15ex}=\hspace{-0.15ex}
u^J_Io_J$, $\mathrm{sdet}\|u^J_I\|\hspace{-0.2ex}\neq
\hspace{-0.2ex}0$ ($\tilde{\gamma} \hspace{-0.15ex}\notin
\hspace{-0.15ex}\{\tilde{o}_I\}$), such that only ${\tilde{t}}'_0,
{\tilde{l}}'_0$ are changed, ${{t}}_0\hspace{-0.15ex}
=\hspace{-0.15ex} -i\tilde{\gamma}^\mu D_\mu, l_0
\hspace{-0.15ex}=\hspace{-0.15ex} -t_0^2$, having obtained a
modified HS symmetry superalgebra,
$\mathcal{A}_{mod}(Y(2),AdS_d)$, b) to construct its auxiliary
representation, the Verma module, by using a Cartan-like
decomposition, extended from the one for the Lie superalgebra
$\{{o}'_I\}\hspace{-0.1ex}\setminus\hspace{-0.1ex}
\{l'_i,l^{\prime+}_i,t'_0,l_0'\}$,
\begin{equation}\label{Cartandecomp}
    \mathcal{A}_{mod}(Y(2),AdS_d) =  \{\{t^{\prime+}_i, l^{\prime ij+},
t^{\prime+}; l^{\prime i+}\} \oplus \{g_0^{\prime i}; t_0', l_0'\}
\oplus \{t'_i,l^{\prime ij}, t';l^{\prime i}\} \equiv
\mathcal{E}^-\oplus H \oplus\mathcal{E}^+,
\end{equation}
c) to realize the Verma module as a formal power series
$\sum_{n\geq 0}\sqrt{r}^n\mathcal{P}_n[(a,a^+)_\mathbf{a}]$ in a
Fock space $\mathcal{H}'$ generated by the same number of creation
and annihilation operators as that for the converted 2nd-class
constraints $\{ o'_I(l'_i, l_i^{\prime +}, t', t^{\prime
+})\}\!:\!\!(a,a^+)^{(r)}_\mathbf{a} \longleftrightarrow
f_i,f^{+}_i
 b_{ij}, b_{ij}^+, b_i, b^{+}_i, b, b^+$, $(b_i, b^{+}_i, b, b^+)$ (for a constrained) LF.

A solution of item a) follows from the above requirement on
$\tilde{O}_I$ to be in involution and from a compactly written
multiplication table \cite{adsfermBKR,0711.4489SQS07} for $\{\tilde{o}_i\}$ in
$\mathcal{A}_{mod}(Y(2),AdS_d)$,
\begin{eqnarray}
\vspace{-2ex} [\,o_I',o_J'\}  =
f_{IJ}^Ko_K'-(-1)^{\varepsilon(o_K)\varepsilon(o_M)}f_{IJ}^{KM}o_M'o_K'
  \quad \mathrm{if} \quad [\tilde{o}_I,\tilde{o}_J\} =
f_{IJ}^K\tilde{o}_K+f_{IJ}^{KM}\tilde{o}_K\tilde{o}_M,
\vspace{-2ex} \label{addal}
\end{eqnarray}
with structure constants $f_{IJ}^K, f_{IJ}^{KM} = -
(-1)^{\varepsilon(o_I)\varepsilon(o_J)}(f_{JI}^K, f_{JI}^{KM})$
and the Grassmann parity $\varepsilon(o_I) = 0,1$, respectively,
for bosonic and fermionic $o_I$. In its turn, the solution of
items b), c) is more involved than the one presented for
$\mathcal{A}'(Y(1),AdS_d)$ \cite{adsfermBKR,0711.4489SQS07} and
 $\mathcal{A}'(Y(2),\mathbf{R}^{1,d-1})$ \cite{flatfermmix},
due to a nontrivial entanglement of a triple $\bigl({l_1^{\prime
+}}t^{\prime +}{l_2^{\prime +}}\bigr)$, being effectively solved
iteratively, thus extending the known results on the Verma module
construction \cite{0206027} and its Fock-space realization in
$\mathcal{H}'$. Note that, within the conversion, we have
$\tilde{M} = \tilde{m} + \tilde{m}' =0$, whereas new
 constants $m_0, h^i$ [they are to be determined later from
the  condition  of reproducing the correct form of
Eqs.(\ref{t't0})] and operators $o'_I$ are found explicitly in
terms of
 $(a,a^+)_\mathbf{a}$ [$(a,a^+)_\mathbf{a}^r$ for constrained LF], as in \cite{adsfermBKR,0206027}.
 \newline
\indent A nilpotent BFV-BRST charge ${Q}'$ for an open
superalgebra\nolinebreak{\,}$\mathcal{A}_{\hspace{-0.1em}conv
}\hspace{-0.1em}(\hspace{-0.1em}Y(2),\hspace{-0.1em}AdS_d\hspace{-0.1em})$
of
 $\tilde{O}_I$ in the  case of Weyl ordering
for the quadratic combinations of $\tilde{O}_I$ in the r.h.s. of
$[\tilde{O}_I, \tilde{O}_J\}$ = $F_{IJ}^K(\tilde{O},
o')\tilde{O}_K$ and for the $(\mathcal{C}\mathcal{P})$-ordering of
the ghost coordinates and momenta $\mathcal{C}^I$,
$\mathcal{P}_I$: bosonic $(q_0, p_0)$,  $(q_i, p_i^+)$,
$(q_i^+,p_i)$ and fermionic $(\eta_0, {\cal{}P}_0)$, $(\eta_i^+,
\mathcal{P}_i)$, $(\eta_i, {\cal{}P}_i^+)$, $(\eta_{ij}^+,
{\cal{}P}_{ij})$, $(\eta_{ij}, {\cal{}P}_{ij}^+)$, $(\eta,
\mathcal{P}^+)$, $(\eta^+, \mathcal{P})$, $(\eta^i_G,
{\cal{}P}^i_G)$, with the standard ghost number distribution
$gh(\mathcal{C}^I)$ = $ - gh(\mathcal{P}_I)$ = $1$, providing
$gh({Q}')$ = $1$  corresponds, at least, to a formal second-rank
topological gauge theory,
\begin{equation}\label{generalQ'}
    {Q}'  = {O}_I\mathcal{C}^I + \textstyle\frac{1}{2}
    \mathcal{C}^{I_1}\mathcal{C}^{I_2}F^J_{I_2I_1}\mathcal{P}_J (-1)^{\varepsilon({O}_{I_2} +
    \varepsilon({O}_J)}+\textstyle\frac{1}{6}\mathcal{C}^{I_1}\mathcal{C}^{I_2}\mathcal{C}^{I_3}
    F^{J_2J_1}_{I_3I_2I_1}\mathcal{P}_{J_2}\mathcal{P}_{J_1} +
    \mathcal{O}(\mathcal{C}^4),
\end{equation}
with completely determined functions
$F^{J_2J_1}_{I_3I_2I_1}(\tilde{O}, o')$ resolving the Jacobi
identity for $\tilde{O}_I$. Note that ${Q}'$ is more involved than
that of \cite{adsfermBKR} for $\mathcal{A}_{\hspace{-0.1em}conv
}\hspace{-0.1em}(\hspace{-0.1em}Y(1),\hspace{-0.1em}AdS_d\hspace{-0.1em})$
and coincides with that of \cite{flatfermmix} for $r=0$
($\mathcal{A}_{\hspace{-0.1em}conv
}\hspace{-0.1em}(\hspace{-0.1em}Y(2),\hspace{-0.1em}\mathbf{R}^{1,d-1}\hspace{-0.1em})$).

A covariant extraction of  $G^i_0 = g_0^i + g_0^{\prime i}(h^i)$
from $\{\tilde{O}_I\}$, that serves to pass to the BFV-BRST charge
$Q$ for the 1st-class constraints $\{{O}_\alpha\}$ only, is based
on the condition of the independence of $\mathcal{H}_{tot}$ of
$\eta^i_G$, and on the elimination from ${Q}'$ of the terms
proportional to $\mathcal{P}^i_G, \eta^i_G: \mathcal{K}^i =
(\sigma^i+h^i)$, as in \cite{flatfermmix}:
\begin{eqnarray}
{Q}' \hspace{-0.2ex}  = \hspace{-0.2ex}   {Q} \hspace{-0.15ex}
+\hspace{-0.15ex} \eta^i_G\mathcal{K}^i\hspace{-0.15ex}
+\hspace{-0.15ex} \mathcal{B}^i \mathcal{P}^i_G;\;
\mathcal{K}^i\hspace{-0.2ex} = \hspace{-0.2ex} G^i_0
\hspace{-0.15ex}+
\hspace{-0.15ex}\bigl(q_i^+p_i\hspace{-0.15ex}+\hspace{-0.15ex}\eta_i^{+}\mathcal{P}^{i}
\hspace{-0.15ex}+\hspace{-0.15ex}
\textstyle\sum_{j}(1\hspace{-0.1ex}+\hspace{-0.1ex}\delta_{ij})\eta_{ij}^+\mathcal{P}^{ij}
\hspace{-0.1ex}+\hspace{-0.1ex}(-1)^i\eta^+\mathcal{P}
\hspace{-0.1ex}+\hspace{-0.1ex} h.c.\bigr)\hspace{-0.1ex};
\label{decomposQ'}
\end{eqnarray}
the same applies to the physical vector $|\chi\rangle \in
\mathcal{H}_{tot}$, $|\chi\rangle = |\Phi\rangle +
|\Phi_A\rangle$, $|\Phi_A\rangle_{\{(a,a^+)_\mathbf{a}=
\mathcal{C} = \mathcal{P} = 0\}}$ = $0$, with the use of the
BFV-BRST equation ${Q}'|\chi\rangle = 0$ that determines the physical
states:
\begin{equation}
\label{Qchi} {Q}|\chi\rangle=0, \quad
(\sigma^i+h^i)|\chi\rangle=0, \quad (\varepsilon,
gh)(|\chi\rangle)=(1,0),
\end{equation}where
the two final equations determine the spectrum of generalized spin
values for $|\chi\rangle$.

The presence of a redundant gauge ambiguity in the definition of
an LF allows one \cite{adsfermBKR,flatfermmix} to ex\-pand $Q$ and $|\chi\rangle$ in the powers of the
zero-mode pairs $q_0, p_0, \eta_0, \mathcal{P}_0$ as follows:
\begin{eqnarray} \label{strQ}
 (Q;|\chi\rangle)\hspace{-0.1em}=\hspace{-0.1em}
 \Bigl(q_0\tilde{T}_0\hspace{-0.1em}+\hspace{-0.1em}\eta_0\tilde{L}_0\hspace{-0.1em}
+\hspace{-0.1em}(\eta_i^+q_i\hspace{-0.1em}-\hspace{-0.1em}\eta_iq_i^+)p_0
\hspace{-0.1em}+\hspace{-0.1em}(q_0^2\hspace{-0.1em}-\hspace{-0.1em}
\eta_i^+\eta_i){\cal{}P}_0
\hspace{-0.1em}+\hspace{-0.1em}\Delta{}Q;
\textstyle\sum_{k=0}^{\infty}q_0^k( |\chi_0^k\rangle
\hspace{-0.1em}+\hspace{-0.1em}\eta_0|\chi_1^k\rangle)\Bigr)\hspace{-0.2ex}
.
\end{eqnarray}
As a result, the 1st equation in (\ref{Qchi}) is Lagrangian and
takes the form (together with the action)
\begin{eqnarray}
&& \Delta{}Q|\chi^{0}_{0}\rangle
+\textstyle\frac{1}{2}\bigl\{\tilde{T}_0,\eta_i^+\eta_i\bigr\}
|\chi^{1}_{0}\rangle =0, \qquad \tilde{T}_0|\chi^{0}_{0}\rangle +
\Delta{}Q|\chi^{1}_{0}\rangle =0, \label{EofM2all}\\
&& \mathcal{S} =
\langle\tilde{\chi}^{0}_{0}|K\tilde{T}_0|\chi^{0}_{0}\rangle +
\textstyle\frac{1}{2}\,\langle\tilde{\chi}^{1}_{0}|K\bigl\{
   \tilde{T}_0,\eta_1^+\eta_1\bigr\}|\chi^{1}_{0}\rangle
+ \langle\tilde{\chi}^{0}_{0}|K\Delta{}Q|\chi^{1}_{0}\rangle +
\langle\tilde{\chi}^{1}_{0}|K\Delta{}Q|\chi^{0}_{0}\rangle,
\label{S1}
\end{eqnarray}
where we have used the operator $K = \hat{1}\bigotimes K'\bigotimes
\hat{1}_{gh}$, providing the Hermiticity of $Q$ w.r.t. $\langle\ |\
\rangle$ in $\mathcal{H}_{tot}$  and the reality of ${\cal{}S}$. The
corresponding LF of an HS field with a given value of spin
$\mathbf{s}=(n_1+\frac{1}{2},n_2+\frac{1}{2})$ is a reducible
gauge theory of $L = (n_1+n_2)$-th stage of reducibility, while
the above LF with constraints has a reduced field spectrum, with $L \leq
3$ for any spin.

\vspace{1ex} \textbf{Acknowledgements:} The author thanks the
organizers of the SPMTP'08 Conference (JINR, Dubna, Russia) for
support and hospitality. The work was supported by the RFBR grant,
project No.\  08-02-08602.

\end{document}